\documentclass[pra,showpacs,twocolumn]{revtex4}
\usepackage{hyperref}
\usepackage{graphicx,psfrag}
\usepackage{amsmath}

\newcommand*{\ket}[1]{|#1\rangle}
\newcommand*{\bra}[1]{\langle#1|}

\newcommand*{\im}{{\rm i}}

\newcommand*{\mapstate}[4]{\widetilde{\ket{#1,#2}}^{[#3,#4]}}

\DeclareMathOperator*{\Tr}{Tr}

\begin{document}

\title{Symmetric qubits from cavity states} \date{\today}

\author{Daniel E. Browne}
\author{Martin B. Plenio}
\affiliation{QOLS, Blackett Laboratory, Imperial College, London, SW7 2BW, UK.}

\begin{abstract}
Two-mode cavities can be prepared in quantum states which represent symmetric multi-qubit states. However, the qubits are impossible to address individually and as such cannot be independently measured or otherwise manipulated.
We propose two related schemes to coherently transfer the qubits which the cavity state represents onto individual atoms, so that the qubits can then be processed individually.
In particular, our scheme can be combined with the quantum cloning scheme of Simon and coworkers [C.~Simon~{\em et~al}, Phys. Rev. Lett. {\textbf{84}}, 2993 (2000)] to allow the optimal clones which their scheme produces to be spatially separated and individually utilized.

\end{abstract}

\maketitle

\section{Introduction}

Experiments using optical and microwave cavities have provided a useful testing ground for the foundations of quantum mechanics and for demonstrating basic principles of quantum information processing \cite{harochermp}.
One of the fundamental consequences of the linearity of quantum mechanics is the no-cloning theorem \cite{nocloning,dieksnocloning}, which states that it is impossible to perfectly copy an arbitrary quantum state. However, approximate copying is possible, and universal optimal quantum cloning devices have been proposed which produce the best possible copies of an arbitrary quantum state \cite{cloning1}.
Simon and coworkers have proposed  \cite{stim1,stim2} a scheme to implement optimal quantum cloning in a simple natural way using atoms and a cavity.

Simon {\em et al}'s scheme utilizes stimulated emission from atoms in a high finesse cavity, where the quantum bits are represented in the state of the two polarization modes of the cavity. Excited atoms with Lambda energy level configurations are passed through the cavity where they emit photons via both stimulated emission and spontaneous emission. The net result of these two processes is that the photons produced are copies of the original photons, due to the stimulated emission, but are noisy, due to the spontaneous emission. Simon and coworkers showed that such a process can in fact generate states in the cavity which represent optimal clones of the initial cavity state.

However, represented in the state of the cavity, the clones are not spatially separate two-level subsystems, and can therefore be neither individually addressed nor measured, which restricts their usefulness.
A suggestion is made in  \cite{stim2} to separate these qubits into polarization states of individual photons by using an array of beamsplitters. This, however, is not an appealing practical proposal for several reasons. To ensure a low probability of finding more than a single photon in each output mode, one would need to use a large number of  beamsplitters and many more output modes than the number of cloned photons. Each beamsplitter will introduce some noise to the photonic qubits, reducing the fidelity of the clones, and, in order to localize the photons, one would need to perform a polarization-independent non-demolition measurement of the number of photons in each output mode.

In this paper we propose two simpler schemes which allow the qubits represented in the cavity state to be transferred coherently onto  spatially separated qubits, which are represented in the energy levels of single atoms.

This process could then be immediately applied to the cavity states produced in Simon {\em et al}'s cloning scheme. This would produce truly spatially separated clones which could be individually addressed.

\begin{figure}
\setlength{\unitlength}{1cm}
\begin{picture}(8,2)

\put(1.8,1.5){\line(1,0){1.6}}
\put(4.6,1.5){\line(1,0){1.6}}
\put(3.2,0.2){\line(1,0){1.6}}
\put(3.6,1.4){$\ket{0}$}
\put(6.4,1.4){$\ket{1}$}
\put(5.0,0.1){$\ket{g}$}
\put(4.4,0.35){\vector(1,1){1.0}}
\put(5.4,1.35){\vector(-1,-1){1.0}}
\put(3.6,0.35){\vector(-1,1){1.0}}
\put(2.6,1.35){\vector(+1,-1){1.0}}

\end{picture}

\caption{\label{vatomfig}The level structure of the three-level atoms utilized in this scheme. The levels are in a V-configuration with two degenerate excited levels $\ket{0}$ and $\ket{1}$ which form the basis for logical qubits, and the ground state $\ket{g}$. The transitions between the two excited levels and the ground state each couple to a different cavity polarization mode.}
\end{figure}
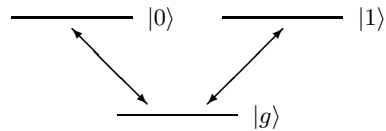

\section{Qubit transfer schemes}\label{mainsection}

In this section we will describe two related schemes to transfer qubits represented by a cavity state onto atoms. This first scheme is deterministic, i.e., every atom which passes through the cavity, will, upon leaving the cavity, have absorbed a photon and had a qubit transferred onto it. The second scheme, on the other hand, depends on the outcome of projective energy measurements on the atoms once they have left the cavity, the successful transfer of a qubit onto the atom occurs, then, in general, with less than unit probability, so this scheme is non-deterministic. First, however, we will briefly discuss the way in which symmetric multi-qubit states can be represented in atoms and a two-mode cavity.

A symmetric multi-qubit state is a state which is invariant under the interchange of any two qubits. The symmetric subspace of a $n$-qubit Hilbert space is spanned by the states $\ket{S(j,n-j)}$ for $j\in\{0,\ldots,n\}$, which are the pure symmetric states with $j$ qubits in state $\ket{0}$ and $(n-j)$ qubits in state $\ket{1}$. These states are defined and discussed further in the appendix.

Consider a cavity with two modes of orthogonal polarization, which we shall label $0$ and $1$. Quantum states of the cavity can be expressed in the Fock-basis of the two modes. If the total number of photons in the cavity is definite, there is a natural correspondence between $n$-photon cavity states and symmetric $n$-qubit states. Let us consider the cavity Fock state $\ket{j,n-j}$, the state where modes $0$ and $1$ contain $j$ and $n-j$ photons respectively, as representing the $n$-qubit symmetric state $\ket{S(j,n-j)}$. Since all basis states of the symmetric $n$-qubit subspace can be represented, any arbitrary symmetric $n$-qubit state can be represented in the cavity.

Additionally, qubits can be represented in the energy levels of atoms. In this paper, we consider three-level atoms with a V-configuration of energy levels, that is, with two degenerate excited states and a single ground state, as illustrated in figure \ref{vatomfig}. We will use the excited subspace of the atoms to represent qubits, and thus label the excited states $\ket{0}$ and $\ket{1}$ to make this clear. The ground state is labeled $\ket{g}$. We will thus consider the atom to be representing a qubit only when its state is wholly within this excited subspace.

In the appendix, in equation  \eqref{decomp}, a decomposition of a general symmetric state in terms of symmetric states on qubit subsets is presented. This means that symmetric states can be represented in a system consisting of a two-mode cavity and a number of atoms, such that each atom carries a qubit and the remainder are represented in the cavity. Let us use the notation, 
$\mapstate{j}{n-j}{n_c}{n_a}$, where $n_c+n_a=n$,
 to refer to the physical state of a cavity and  $n_a$ atoms which represents the symmetric state $\ket{S(j,n-j)}$,  with $n_a$ qubits each represented in the excited subspace of $n_a$ atoms and the remaining $n_c$ qubits in the cavity state.

Such a state has the following physical representation which we will prove below,

\begin{equation}
\widetilde{\ket{j,n-j}}^{[n_c,n_a]}=\left(\prod_{k=1}^{n_a}\frac{\ket{0}^{A_k} a_0 + \ket{1}^{A_k} a_1}{\sqrt{n-(k-1)}}\right)\ket{j,n-j}
\end{equation}
where the atoms are labeled $A_k$, for $k\in\{1,\ldots,n_a\}$ and where $a_0$ and $a_1$ are annihilation operators for modes $0$ and $1$ respectively.

\subsection{Deterministic Scheme}

The principle of the deterministic scheme is very simple. Atoms, with a V-configuration of energy levels, are passed, in their ground state, through the cavity, one at a time, such that each remains in the cavity for a half integer number of Rabi-oscillations. The atoms will emerge from the cavity in an excited state, and the quantum state of the whole system still represents the same $n$-qubit symmetric state as the initial cavity state, but now, with one of the qubits represented in the excited subspace of the atom.

While in the cavity field the atoms interact with the cavity modes according to the following interaction Hamiltonian. The transitions between the excited states $\ket{0}$ and $\ket{1}$ and the ground state $\ket{g}$ each couple to a different cavity mode with the same coupling constant $\gamma$ for each transition.

\begin{equation}\label{intham}
H_{i}=\hbar\gamma\left(a_0 \ket{0}\bra{g} + a_1 \ket{1}\bra{g} +a_0^\dag \ket{g}\bra{0}+a_1^\dag \ket{g}\bra{1}\right)
\end{equation}
where $a_0$ and $a_1$ are annihilation operators for modes $0$ and $1$ respectively.

Let us first consider the simple case where the cavity state is initially in a pure Fock state $\ket{j,n-j}$ representing the $n$-qubit symmetric state $\ket{S(j,n-j)}$. A atom, with a V-configuration of energy levels, as described above, enters the cavity in its ground state and the initial state of the system $\ket{\psi_0}$ is thus
\begin{equation}
\ket{\psi_0}=\ket{g}\ket{j,n-j}=\ket{g}\mapstate{j}{n-j}{n}{0}
\end{equation}

Evolving under $H_i$, the state of the system  undergoes Rabi oscillations
\begin{equation}\label{expev}
\begin{split}
\ket{\psi(t)}&=e^{-\im H_{i} t/\hbar} \ket{\psi_0}\\
&=\cos(\sqrt{n}\gamma t) \ket{\psi_0}- \im \sin(\sqrt{n}\gamma t) \ket{\psi_1}
\end{split}
\end{equation}

where $\ket{\psi_1}=\frac{H_i}{\hbar \sqrt{n}\gamma}\ket{g}\ket{j,n-j}$.
Let us consider the properties of $\ket{\psi_1}$,

\begin{equation}\label{psi1}
\begin{split}
\ket{\psi_1}&=\frac{H_i}{\hbar \sqrt{n}\gamma}\ket{g}\ket{j,n-j}\\&
=\frac{1}{\sqrt{n}}\Bigl(\ket{0}^{A_1} a_0 + \ket{1}^{A_1} a_1\Bigr)\ket{j,n-j}\\
&=\frac{1}{\sqrt{n}}\Bigl(\sqrt{j}\ket{0}^{A_1}\ket{j-1,n-j}\\&\quad+\sqrt{n-j}\ket{1}^{A_1}\ket{j,n-j-1}\Bigr).
\end{split}
\end{equation}

The symmetric $n$-qubit state which $\ket{j,n-j}$ represents, $\ket{S(j,n-j)}$, can be decomposed according to equation \eqref{decomp} in the following way

\begin{equation}
\begin{split}
\ket{S(j,n-j)}=&\frac{1}{\sqrt{n}}\Bigl(\sqrt{j}\ket{S(1,0)}\otimes\ket{S(j-1,n-j)}\\&+\sqrt{n-j}\ket{S(0,1)}\otimes\ket{S(j,n-j-1)}\Bigr).
\end{split}
\end{equation}

However, $\ket{S(1,0)}=\ket{0}$ and $\ket{S(0,1)}=\ket{1}$. Comparing this with equation \eqref{psi1} we see that $\ket{\psi_1}$ represents the symmetric state $\ket{S(j,n-j)}$ with one of the qubits represented on atom $A_1$, and the cavity state representing the remaining $(n-1)$ qubits. Thus $\ket{\psi_1}=\mapstate{j}{n-j}{n-1}{1}$.
Thus, we see that if the atom remains in the cavity for a half integer number of Rabi periods, then one of the qubits of the $n$-qubit state which was represented in the cavity will have been transferred onto the atom.

We can show inductively, that if this is repeated with further atoms, each time, a further qubit is transferred to each atom.
Consider the system of a cavity and $m$ atoms, which represents the symmetric state $\ket{S(j,n-j)}$ with $m$ qubits represented on the atoms and the remainder in the cavity, i.e., the system is in 
 the state $\mapstate{j}{n-j}{n-m}{m}$. A further atom, which we will label,  $A_{(m+1)}$ is introduced to the cavity and they interact under $H_i$ in the same way as described above, except that the Rabi frequency is now $2(\sqrt{n-m})\gamma$ as only $(n-m)$ photons remain in the cavity. Half way through the Rabi oscillation cycle, the system is in the state $\ket{\psi_{m+1}}$,

\begin{equation}
\begin{split}
\ket{\psi_{m+1}}&=\frac{H_i}{\hbar \sqrt{n-m}\gamma}\ket{g}\mapstate{j}{n-j}{n-m}{m}\\
&=\frac{1}{\sqrt{n-m}}\Bigl(\ket{0}^{A_{(m+1)}} a_0 \\&\quad+ \ket{1}^{A_{(m+1)}} a_1\Bigr)\mapstate{j}{n-j}{n-m}{m}
\end{split}
\end{equation}
where the operator $H_i$  acts upon atom $A_{m+1}$ and the cavity modes.

We need to know how the cavity mode annihilation operators $a_0$ and $a_1$ act on $\mapstate{j}{n-j}{n-m}{m}$. We can decompose state $\mapstate{j}{n-j}{n-m}{m}$ using equation \eqref{decomp} in terms of symmetric states on $m$ atoms and $(n-m)$-photon cavity states

\begin{equation}
\begin{split}
&\mapstate{j}{n-j}{n-m}{m}\\&\qquad=\sqrt{\frac{1}{{n \choose j}}}\sum_{k=0}^{m} \sqrt{{m \choose k}{n-m \choose j-k}} \mapstate{k}{m-k}{0}{m}\\&\qquad\quad\otimes\mapstate{j-k}{(n-m)-(j-k)}{n-m}{0}.
\end{split}
\end{equation}

Since $\mapstate{j}{n-j}{n}{0}=\ket{j,n-j}$, we can thus calculate  $a_0\mapstate{j}{n-j}{n-m}{m}$ and $a_1\mapstate{j}{n-j}{n-m}{m}$. For example,

\begin{equation}
\begin{split}
&a_0\mapstate{j}{n-j}{n-m}{m}\\&\qquad=\sqrt{\frac{1}{{n \choose j}}}\sum_{k=0}^{m} \sqrt{{m \choose k}{n-m \choose j-k}} \mapstate{k}{m-k}{0}{m}\\&\qquad\quad\otimes \sqrt{j-k}\ket{j-k-1,(n-m)-(j-k)}.
\end{split}
\end{equation}
Using the fact that $(j-k){n-m \choose j-k}=(n-m) {n-m-1 \choose j-k-1}$ and ${n \choose j}=(n/j){n-1 \choose j-1}$ this simplifies to

\begin{equation}\begin{split}
&a_0\mapstate{j}{n-j}{n-m}{m}\\&\qquad=\sqrt{\frac{(n-m)j}{n}}\mapstate{j-1}{n-j}{n-m-1}{m}.\end{split}
\end{equation}

Similarly one finds

\begin{equation}
\begin{split}
&a_1\mapstate{j}{n-j}{n-m}{m}\\&\qquad=\sqrt{\frac{(n-m)(n-j)}{n}}\mapstate{j}{n-j-1}{n-m-1}{m}.
\end{split}
\end{equation}

We can use this to calculate $\ket{\psi_{m+1}}$

\begin{equation}
\begin{split}
&\ket{\psi_{m+1}}\\&\quad=\frac{1}{\sqrt{n-m}}(\ket{0}^{A_{(m+1)}} a_0 + \ket{1}^{A_{(m+1)}} a_1)\mapstate{j}{n-j}{n-m}{m}\\
&\quad=\sqrt{\frac{1}{n}}\Bigl(\sqrt{j}\ket{0}^{A_{(m+1)}}\mapstate{j-1}{n-j}{n-m-1}{m}\\&\quad\qquad+
\sqrt{n-j}\ket{1}^{A_{(m+1)}}\mapstate{j}{n-j-1}{n-m-1}{m}\Bigr)\\
&\quad=\mapstate{j}{n-j}{n-(m+1)}{(m+1)}.
\end{split}
\end{equation}

Thus, after a $m$ qubits have already  been transferred from the cavity onto atoms, passing a further atom through the cavity, for the appropriate duration, does indeed transfer a further qubit from the cavity onto this atom. We can then use induction to find a general way of expressing  $\widetilde{\ket{j,n-j}}^{[n_c,n_a]}$ in terms of physical states and operators. Since $\mapstate{j}{n-j}{n}{0}=\ket{j,n-j}$,

\begin{equation}
\widetilde{\ket{j,n-j}}^{[n_c,n_a]}=\left(\prod_{k=1}^{n_a}\frac{\ket{0}^{A_k} a_0 + \ket{1}^{A_k} a_1}{\sqrt{n-(k-1)}}\right)\ket{j,n-j}.
\end{equation}

The scheme also works if the cavity is prepared in a mixed state where all the terms in the mixture have the same total photon number $n$. Such a state will represent a mixed $n$-qubit symmetric state, i.e., the cavity state
\begin{equation}\label{rhocav}
\rho_{c}=\sum_{j=0}^n\sum_{k=0}^n c_{j,n-j;k,n-k}\ket{j,n-j}\bra{k,n-k}
\end{equation}
represents the symmetric state $\rho_s$
\begin{equation}
\rho_{s}=\sum_{j=0}^n\sum_{k=0}^n c_{j,n-j;k,n-k}\ket{S(j,n-j)}\bra{S(k,n-k)}.
\end{equation}

When an atom enters the cavity, because all terms have the same total photon number, the evolution of the system is a simple Rabi oscillation and after half a Rabi period, at time $t=(\pi/2\sqrt{n}\gamma)$, the system is in state $\rho_1$
\begin{equation}
\begin{split}
\rho_1=&\frac{H_i}{\sqrt{n}\hbar\gamma}\rho_c\otimes\ket{g}\bra{g}\frac{H_i}{\sqrt{n}\hbar\gamma}\\
&=\sum_{j=0}^n\sum_{k=0}^n c_{j,n-j;k,n-k}\\&\qquad\times\frac{H_i}{\sqrt{n}\hbar\gamma}\ket{j,n-j}\ket{g}\bra{g}\bra{k,n-j}\frac{H_i}{\sqrt{n}\hbar\gamma}
\end{split}
\end{equation}
Every term in the mixture, is acted upon from both sides by ${H_i}{\sqrt{n}\hbar\gamma}$, which we have shown above to have the effect of transferring a qubit from the cavity onto the atom. Thus, as desired,

\begin{equation}\begin{split}
\rho_1&=\sum_{k=0}^n c_{j,n-j;k,n-k}\mapstate{j}{n-j}{n-1}{1}\widetilde{\bra{j,n-j}}^{[n-1,1]}\\&=\tilde \rho^{[n-1,1]}
\end{split}\end{equation}
where we introduce a notation for mixed states analogous to that for pure states, i.e., $\tilde \rho^{[n_c,n_a]}$ is the physical state which represents the mixed $(n_a+n_c)$-qubit state $\rho$ such that $n_c$ qubits are represented in the state of the cavity and $n_a$ qubits on $n_a$ atoms.

In the same way as shown above for Fock states, one can show inductively that repeating this with further atoms, such that each remains in the cavity for a half integer multiple of the appropriate Rabi period, will, as for the Fock states, transfer further qubits from the cavity onto the atoms until the cavity has reached the vacuum state and all the qubits of the state are represented in individual atoms. One finds the following expression for $\tilde \rho^{[n-m,m]}$,

\begin{equation}\begin{split}
\tilde \rho_n^{[n-m,m]}&=\left(\prod_{k=1}^{m}\frac{H^{(k)}_i}{\hbar \sqrt{n-(k-1)} \gamma}\right)\rho_n\biggl(\ket{g}\\&\qquad\bra{g}\biggr)^{\otimes m}\left(\prod_{k'=1}^{m}\frac{H^{(k')}_i}{\hbar \sqrt{n-(k'-1)} \gamma}\right).
\end{split}\end{equation}
where  $H_i^{(k)}$ is the interaction Hamiltonian $H_i$ between the cavity and  the $k$-th atom.

\subsection{Non-deterministic Scheme}

To implement the scheme described in the previous section it is necessary to know, at each step in the process, the total photon number of the cavity state, in order to calculate the length of time which each atom should interact with the cavity. However, the photon number of the  cavity may not be known or the cavity state may be a mixture of states with different total photon numbers.
For example, the states created by  Simon {\em et al}'s general cloning scheme are mixtures of states representing different numbers of optimally cloned qubits. A  non demolition measurement of the total number of photons in the cavity or a measurement of the energies of the atoms used in the cloning process will project the cavity state into one of these states. Otherwise, the cavity remains in a mixed state.

We can adapt the scheme described in the previous section for these situations by adding a further element to the process. If, upon leaving the cavity, the energy of each atom is measured, this measurement will project the system into one of two outcomes. Either the atom is in its ground state and no qubit transfer took place, or the atom has been excited and the qubit transfer was successful. This works, even if the interaction time is not a half integer multiple of the Rabi period, different interaction times merely change the probability of a successful qubit transfer and make the scheme non-deterministic.

To see this, let us first consider the simple case of a cavity in the pure Fock state of the cavity $\ket{j,n-j}$ and a single atom. The atom, in the ground state of its $V$-configuration of energy levels, enters the cavity and interacts with the cavity modes via the interaction $H_i$ as described above. A Rabi oscillation occurs and if after some time $\tau$ the system is in general in an entangled superposition of states,

\begin{equation}\begin{split}
\ket{\psi(\tau)}=&\cos(\sqrt{n}\gamma \tau)\mapstate{j}{n-j}{n}{0}\ket{g}\\&-\im \sin(\sqrt{n}\gamma \tau)\mapstate{j}{n-j}{n-1}{1}.
\end{split}\end{equation}

If the energy of the atom is now measured, the system will be projected into $\mapstate{j}{n-j}{n}{0}\ket{g}$ where no qubit transfer has occurred if the ground state is measured, or the state $\mapstate{j}{n-j}{n-1}{1}$ where one qubit has been successfully transferred to the qubit, if the atom is measured to be excited.
The probability that a successful transfer of a qubit onto the atom will occur is $\sin^2(\sqrt{n}\gamma \tau)$.
Since the two excited  levels $\ket{0}$ and $\ket{1}$ are degenerate, an energy measurement will not distinguish these two states, thus a qubit state in the ($\ket{0}$,$\ket{1}$) subspace of the atom will be undisturbed by a projective energy measurement.

As long as one accepts a smaller probability of success, in contrast to the deterministic scheme, the interaction time need not be finely tuned to a half integer multiple of the Rabi period.
This allows one of the inconvenient aspects of the deterministic scheme to be avoided, namely the need to have a different interaction time for each atom passing through the cavity, with the trade off that the qubit transfer will not be successful each time. 

One can similarly show that this scheme is also effective for general mixed states with a fixed total photon number, in the same way as for the deterministic scheme, since the Rabi frequency of the interaction for all terms in the mixture is the same.
This non-deterministic scheme, however, is successful even if the cavity is in a mixture of states with different photon numbers, representing a mixture of different symmetric $n$-qubit states for different values of $n$, like the states produced in Simon {\em et al}'s cloning scheme.  

\begin{figure}
\includegraphics[scale=1.0]{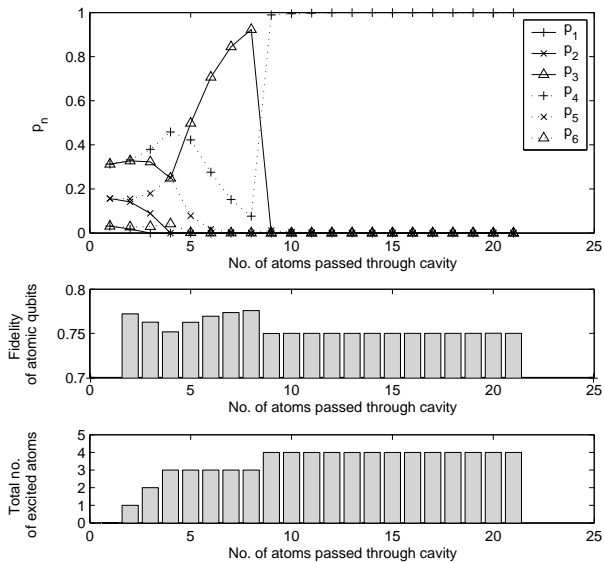}
\caption{These graphs show the results of a numerical simulation of the non-deterministic scheme. The
 upper graph shows how the weight of the $n$-qubit states $p_n$ evolve as atoms are passed, one-by-one, through the cavity and their energies measured. 
The initial state of the cavity was a mixture of states representing different numbers of optimal clones of an initial qubit. At each stage of the process, therefore, the atoms which carry qubits are themselves clones of this state, and their fidelity at each stage, calculated using equation~\eqref{atomfid}, is shown in the central graph.
The number of excited atoms which have been detected, and thus the number of qubits which have been transferred to atoms, is illustrated, for each step, in the lower graph. The initial probability distribution is binomial, with a maximum photon number of 6 and a minimum of 1. The interaction duration $\tau$ is set to optimal value for these initial probabilities, $\tau=0.825/\gamma$.}\label{bingraph}
\end{figure}
Consider a cavity in a mixture of states $\rho_n$, each with a different total photon number $n$ and corresponding weight $p_n$, and a V-configuration atom prepared in its ground state. The state of the system at this time $t=0$ is

\begin{equation}
\rho(0)=\sum_n p_n \rho_n\ket{g}\bra{g}.
\end{equation}

The atom enters the cavity at this time and they interact. The evolution of the system is governed by the unitary operator $U(t)=\exp[-\im (H_i/\hbar)t]$. Thus $\rho(t)$ is
\begin{equation}
\begin{split}
\rho(t)&=U(t)\rho(0)U^\dag(t)\\
&=\sum_n p_n\Bigl[U(t)\rho_n \ket{g}\bra{g} U^\dag(t)\Bigr].
\end{split}
\end{equation}

Each term in the mixture evolves as described in the previous section, oscillating with a  Rabi frequency $2\sqrt{n}\gamma$  dependent upon $n$, the total photon number of the term, between the initial state $\rho_n\otimes\ket{g}\bra{g}$, and the state $\tilde \rho_n^{[n-1,1]}$.
When the atom exits the cavity at time $t=\tau$ these oscillations cease, in general, each term will be in a different point of the oscillatory cycle.

The atom's energy is then measured and with probability $p_g=\sum_n p_n\cos^2(\sqrt{n}\gamma \tau )$ the atom is found to be in its ground state. This means that no photon was absorbed and no qubit has been transferred onto the atom. Since the terms in the mixture were at different points in their oscillation, however, the system is not projected back into its initial state, but the weights in the mixture change, reflecting the fact that the atomic measurement reveals a small amount of information about the total photon number in the cavity. The weights $p_n$ change according to,

\begin{equation}
p_n\to p_n\frac{\cos^2(\sqrt{n}\gamma \tau )}{\sum_n p_n\cos^2(\sqrt{n}\gamma \tau )}.
\end{equation}

Alternatively, with probability $p_e=\sum_n p_n \sin^2(\sqrt{n}\gamma\tau)=1-p_g$, the measurement reveals that the atom is excited. This means that a photon has been absorbed by the atom. The system is now projected into a mixture of states, each corresponding to the same $n$-qubit states as before, but with one qubit carried by the atom

\begin{equation}
\rho=\sum_np_n\tilde \rho_n^{[n-1,1]}
\end{equation}
and with changed weights as follows

\begin{equation}
p_n\to p_n\frac{\sin^2(\sqrt{n}\gamma \tau )}{\sum_n p_n\sin^2(\sqrt{n}\gamma \tau )}.
\end{equation}

This state can be interpreted in the following way. Each term in the mixture represents a $n$-qubit state of which one qubit is carried on the atom's excited subspace. The weights of the mixture have changed to reflect the fact that the likelihood of the atom absorbing the photon with a particular interaction time is different for different photon numbers, thus the measurement outcome gives some information about the total number of the cavity, and the weights of the terms change accordingly.

This process is repeated with further atoms, and each time, either a further qubit is transferred to the atom, or the transfer is unsuccessful. The weights of the terms in the mixture change after each repetition.
An example of this is illustrated in figure \ref{bingraph} for an initial cavity state which is a mixture of states with photon numbers from 1 to 6 with a binomial probability distribution, $p_n={6-1\choose n-1}/2^{5}$.

Each time, the probability that a successful qubit transfer will occur is $p_{e}=\sum_n p_n \sin^2(\sqrt{n}\gamma\tau)$. If the probabilities $p_n$ are known, and the duration of the interaction $\tau$  can be controlled, this can chosen so that $p_e$ is maximized. To carry out the process with maximal efficiency, this optimal interaction time can be recalculated after each measurement to reflect the new weights in the mixture.

Eventually, no excited atoms are detected for many repetitions. This can mean either that with high probability the cavity is now in its vacuum state or, if the interaction time between atoms and cavity is being kept constant for each atom, that the system has reached a ``trapping state''  \cite{weidinger}. This is not actually a quantum state of the system, but describes the situation when the interaction time between atom and cavity is very close to a multiple of the interaction's  Rabi period. This means that when the energy of the atom is measured on leaving the cavity, the probability of it being excited and correspondingly the probability of a successful qubit transfer is very low.

The condition for a trapping state is $\gamma\tau=\pi m / \sqrt{n}$, where $m$ is a natural number and $n$ is the total photon number of the cavity.
This trapping condition can coincide for different total photon numbers, for example, for all the square numbers $n=1,4,9,\ldots$

\begin{figure}
\psfrag{amean}{$a_{\text{mean}}$}
\psfrag{sigmarel}{$\sigma_{\text{rel}}$}
\includegraphics[scale=1.0]{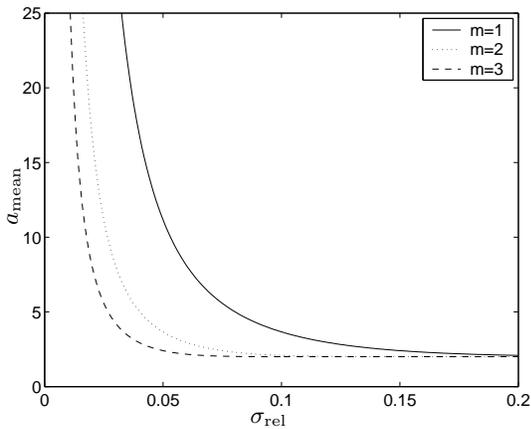}
\caption{This graph plots $a_{\text{mean}}$, the mean number of atoms which have passed through the cavity before the system leaves the ``trapping state'', against the relative standard deviation in the interaction time, $\sigma_{\text{rel}}$ for $m=1,2,3$. }\label{trapgraph}
\end{figure}

Clearly, the trapping phenomenon is only evident when the interaction time for each atom is the same. In practice, however, even if the mean velocity of the atoms is constant there will be a finite velocity distribution. We can show  that this leads to a finite probability of successful qubit transfer, even if the mean velocity fulfills the trapping condition, and that this probability grows with the spread of the velocity distribution.

Let the state of the cavity be some arbitrary mixed state with a definite total photon number $n$. Atoms pass through the atom one at a time with interaction time $\tau$. Let us consider the case where $\tau$ has a Gaussian probability distribution $p(\tau)$ with mean time $\tau_0$ and standard deviation $\sigma$.
The mean interaction time fulfills the trapping state condition $\tau_0 =(\pi m )/(\gamma \sqrt{n})$, where $n$ is the total photon number of the cavity and $m$ is a natural number corresponding to the number of Rabi oscillations performed while the atom is in the cavity.

When the energy of an atom is measured, on leaving the cavity, the probability of a successful qubit transfer is $\sin^2(\sqrt{n}\gamma\tau)$, we can therefore calculate the mean probability of a successful qubit transfer for a Gaussian distribution of $\tau$

\begin{equation}\begin{split}
\bar p(n,\sigma)&=\int_{-\infty}^{+\infty}p(\tau)\sin^2(\sqrt{n}\gamma\tau)d\tau\\&=\frac{1}{\sigma
\sqrt{2\pi}}\int_{-\infty}^{+\infty} e^{-\frac{(\tau-\tau_0)^2}{2\sigma^2}}\sin^2(\sqrt{n}\gamma\tau)d\tau
\end{split}\end{equation}
Let us use the substitution $\tau'=\tau-\tau_0$ and the fact that $\sin^2(\sqrt{n}\gamma\tau)=\sin^2(\sqrt{n}\gamma\tau')$
\begin{equation}
  \bar p(n,\sigma)=\frac{1}{\sigma\sqrt{2\pi}}\int_{-\infty}^{+\infty} e^{-\frac{(\tau')^2}{2\sigma^2}}\sin^2(\sqrt{n}\gamma\tau')d\tau'
\end{equation}
This definite integral can be written in the closed form,
\begin{equation}
\bar p(n,\sigma)=\frac{1}{2}\left(1-e^{-2n\gamma^2\sigma^2}\right)
\end{equation}

The mean number of atoms $a_{\text{mean}}$ which must pass through the cavity, before  a photon is absorbed and the system escapes from the trapping state is simply the inverse of this mean probability

\begin{equation}
a_{\text{mean}}=\frac{1}{\bar p(n,\sigma)}=\frac{2}{1-e^{-2n\gamma^2\sigma^2}}.
\end{equation}
Physically, the relative standard deviation $\sigma_{\text{rel}}=\sigma/\tau_0$ is a more meaningful expression of the uncertainty in $\tau$ than the absolute value $\sigma$, we therefore recast $a_{\text{mean}}$ in terms of $\sigma_{\text{rel}}$ and see that it is independent of $n$.

\begin{equation}
a_{\text{mean}}=\frac{2}{1-e^{-8\pi^2m^2\sigma_{\text{rel}}^2}}.
\end{equation}

We plot this for $m=1,2,3$ in figure \ref{trapgraph}. The higher the value of $m$, the more rapidly $a_{\text{mean}}$ tends towards the limit 2, which is the value of $a_{\text{mean}}$ obtained for a uniform probability distribution in $\tau$.
 This means that trapping states for high values of $m$ need only a small uncertainly in the velocity for them to be easy to escape after a small number of repetitions. The trapping states with $m=1$ are the most robust against uncertainty in the interaction time, but a reasonably large uncertainty is enough to reduce the number of repetitions needed to escape the trapping state to an acceptable value. For example, if $\sigma_{\text{rel}}$ is 0.06, the average number of repetitions needed to escape a $m=1$ trapping state is close to 8.

If the maximum total photon number of the initial state is known, as is the case for the states generated in the optimal cloning scheme, trapping states can be completely avoided if an appropriate interaction time is chosen. If $\tau<(\pi)/(\gamma\sqrt{n_{\text{max}}})$ then all trapping states for values of $n$ up to $n_{\text{max}}$ are avoided.

\begin{figure}
\psfrag{nmax=10pad}{{$n_{\text{max}}=10$}}
\psfrag{nmax=20pad}{$n_{\text{max}}=20$}
\psfrag{nmax=40pad}{$n_{\text{max}}=40$}
\includegraphics[scale=1.0]{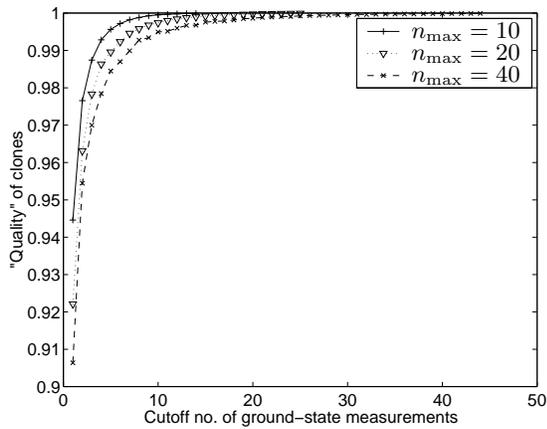}
\caption{The non-deterministic scheme was simulated numerically. This graph shows the average quality of the atomic qubits produced plotted against the number of consecutive  ground state measurements, after which the scheme is stopped.
The initial cavity states are mixtures of states with total photon number $n$ between $n=1$ and $n=n_{\text{max}}$ for $n_{\text{max}}=10,20$ and $40$. Each state represents the output of an optimal $1\to n$ cloner.
Each data-point shows the average quality taken over 1000 runs of the simulation for a particular cutoff number of ground state measurements.
The interaction times used were the optimum for each initial state, but were held constant for the whole process.
}\label{cutoff}
\end{figure}

As described above, Simon {\em et al}'s cloning scheme generates a mixture of states in the cavity, each of which represents the optimal state of $m$ clones from $n$ original qubits in the state $\ket{\psi_{in}}$ for different values of $m$. The fidelity of the clones in each of these states is a function of $n$ and $m$  \cite{bruss98b,keylwerner}.
\begin{equation}
F_{n\to m}=\frac{nm +n + m}{m(n+2)}
\end{equation} 
Thus the reduced state of a single atomic qubit is a mixture of clones of different fidelities, and is therefore itself a clone of $\ket{\psi_{in}}$. Its fidelity $F_{\text{atom}}$ is simply the average fidelity of the terms in the mixture, due to the linearity of the trace operation.
\begin{equation}\label{atomfid}
\begin{split}
F_{\text{atom}}&=\bra{\psi_{in}}\rho_{atom}\ket{\psi_{in}}\\&=\bra{\psi_{in}}\Tr_{\text{all other sub-systems}}\left[\sum_m (p_m \rho_m)\right]\ket{\psi_{in}}\\&=\sum_m p_m F_{n\to m}
\end{split}
\end{equation}

This means that as the scheme progresses, the fidelity of the clones on the atoms, changes as the weights in the mixture change, and can be higher, at an early stage in the process, when fewer qubits have been transferred to atoms, than the final fidelity or even the optimal fidelity for the final number of qubits which are transferred.
Although this seems rather counterintuitive, this can be explained by remembering that the fidelity is a property of an ensemble of systems, and thus reflects an average of the possible final fidelities, which can indeed be higher than the actual final fidelity.

In the limit of large numbers of repetitions, as long as trapping states have been avoided, one term of the original mixture of different numbers of clones will become entirely dominant, and the weights of other terms will become vanishingly small. All of the qubits represented by this term will have been transferred onto atoms, and their states will, in the limit of many atoms, converge to the optimal fidelity. 

In practice however, one cannot send infinitely many atoms through the cavity. A long series of measurements of the ground state of atoms leaving the cavity  is a good indication that the cavity is empty, as long as trapping states have been avoided. Thus if one chose to stop the process after a cutoff, a certain number of ground state measurements, then the fidelity of the qubits would be less than optimal due to the small possibility that another qubit remained in the cavity. To judge, then, what a suitable cutoff would be let us define the ``quality''  of the clones as the ratio of the clone fidelity of the atomic qubits to the optimal fidelity of that number of clones. Numerical simulations of the scheme have shown that a high quality can reached for reasonably low cutoff numbers. The average achieved quality as a function of cutoff is plotted in figure \ref{cutoff} for different initial probability distributions. It is dependent both upon the interaction time chosen and the probability distribution of the terms in the initial cavity state.

\section{Conclusion}

We have introduced two related schemes which allow one to transfer the symmetric n-qubit state represented in the state of a two-mode cavity onto up to $n$ atomic qubits, such that all qubits are then represented in separate subsystems. This would allow the potentially useful symmetric states which can be generated in the cavity, such as the optimally cloned qubits of Simon {\em et al}'s scheme, to be transferred onto atoms, where they can then be individually addressed.

Both schemes have drawbacks which could cause difficulties in their experimental implementation. The precise control of the timing of the atom-cavity interactions necessary in the deterministic scheme could be realized by the application of an external field across the cavity mirrors. This would allow one to shift the atomic energy levels in and out of resonance with the cavity modes at will, and thus achieve the required interaction times.  \cite{harochermp}.

The projective measurement of the atomic energies could be implemented using quantum jump detection techniques  \cite{qjump1,qjump2,plenioqjump}. This would provide a high detection efficiency and would not destroy the qubit state in the measurement process.

After the atoms have left the cavity, they will still be traveling at high velocity, and it would be difficult to cool and trap them without destroying the qubit state. If, however, a stationary qubit is needed, the atom can be passed through a further cavity of the same dimensions, in its vacuum state. If the atom interacts with the cavity for half the Rabi oscillation period, the qubit is mapped onto the cavity, where it is represented in the ($\ket{1,0}$,$\ket{0,1}$) basis, in a similar way to the ``quantum memory'' scheme implemented in  \cite{maitre97}.

The identification between the states of a two-mode cavity states and symmetric qubit states is natural, but the chosen correspondence between the Fock basis states and the symmetric basis states is not the only one one could make, and could be considered to be an arbitrary choice. These schemes give a concrete physical justification for this choice.

\begin{acknowledgments}
D.E.B. would like to thank Angelo Carollo, Julia Kempe and Christoph Simon for helpful discussions.
This work was supported by the EPSRC, the EQUIP project of the European Union, Hewlett-Packard Ltd. and U.S. Army grant no. DAAD19-02-1-0161.

\end{acknowledgments}

\begin{appendix}
\section{Symmetric States}

In this appendix we will introduce symmetric $n$-qubit states and discuss a useful way in which they can be written in terms of symmetric states on subsets of the qubits.

Let $\ket{S(j,n-j)}^{(1,\ldots,n)}$ be a state of a system of $n$-qubits labeled 1 to $n$, which is completely symmetric under the interchange of any two qubits, such that $j$ qubits are in state $\ket{0}$ and $n-j$ qubits in state $\ket{1}$. It can be shown that this state is unique except for a global phase, which is unimportant for this discussion.

We will thus use the following definition for $\ket{S(j,n-j)}^{(1,\ldots,n)}$ which fulfills the above conditions.

\begin{equation}\begin{split}
&\ket{S(j,n-j)}^{(1,\cdots,n)}\\&\qquad=\sqrt{\frac{1}{{n \choose j}}}\sum_{\text{permutations}} \underbrace{\ket{0}^1\cdots\ket{0}^j}_j\underbrace{\ket{1}^{j+1}\cdots\ket{1}^n}_{n-j},
\end{split}\end{equation}

where the permutations are of the values $0$ and $1$ with respect to the qubit labels.

In our discussion in section \ref{mainsection}, we use the following decomposition of symmetric states.
Let us divide our $n$ qubits into two sets, the first of qubits $1$ to $m$ and the second the remaining qubits $(m+1)$ to $n$ for some integer $m$ between 1 and $n-1$. We can write the symmetric state $\ket{S(j,n-j)}^{(1,\ldots,n)}$ in terms of symmetric states of each of these qubit subsets as follows:

\begin{equation}\label{decomp}\begin{split}
&\ket{S(j,n-j)}^{(1,\cdots,n)}\\&\qquad=\sqrt{\frac{1}{{n \choose j}}}\sum_{k=0}^{m} \sqrt{{m \choose k}{n-m \choose j-k}} \ket{S(k,m-k)}^{(1,\cdots,m)}\\&\qquad\quad\otimes\ket{S(j-k,(n-m)-(j-k))}^{(m+1,\cdots,n)}.
\end{split}\end{equation}

\end{appendix}

%\bibliographystyle{apsrev}
%\bibliography{journals,clonebib,newbib,cavityqed}

\begin{thebibliography}{10}
\expandafter\ifx\csname bibnamefont\endcsname\relax
  \def\bibnamefont#1{#1}\fi
\expandafter\ifx\csname bibfnamefont\endcsname\relax
  \def\bibfnamefont#1{#1}\fi
\expandafter\ifx\csname url\endcsname\relax
  \def\url#1{\texttt{#1}}\fi
\expandafter\ifx\csname urlprefix\endcsname\relax\def\urlprefix{URL }\fi
\providecommand{\bibinfo}[2]{#2}
\providecommand{\eprint}[2][]{\url{#2}}

\bibitem{harochermp}
\bibinfo{author}{\bibfnamefont{J.}~\bibnamefont{Raimond}},
  \bibinfo{author}{\bibfnamefont{M.}~\bibnamefont{Brune}}, \bibnamefont{and}
  \bibinfo{author}{\bibfnamefont{S.}~\bibnamefont{Haroche}},
  \bibinfo{journal}{Rev. Mod. Phys.} \textbf{\bibinfo{volume}{73}},
  \bibinfo{pages}{565} (\bibinfo{year}{2001}).

\bibitem{nocloning}
\bibinfo{author}{\bibfnamefont{W.~K.} \bibnamefont{Wootters}} \bibnamefont{and}
  \bibinfo{author}{\bibfnamefont{W.~H.} \bibnamefont{Zurek}},
  \bibinfo{journal}{Nature (London)}
  \textbf{\bibinfo{volume}{299}}(\bibinfo{number}{5886}), \bibinfo{pages}{802}
  (\bibinfo{year}{1982}).

\bibitem{dieksnocloning}
\bibinfo{author}{\bibfnamefont{D.}~\bibnamefont{Dieks}},
  \bibinfo{journal}{Phys. Lett. A} \textbf{\bibinfo{volume}{92}},
  \bibinfo{pages}{271} (\bibinfo{year}{1982}).

\bibitem{cloning1}
\bibinfo{author}{\bibfnamefont{V.}~\bibnamefont{Bu\v{z}ek}} \bibnamefont{and}
  \bibinfo{author}{\bibfnamefont{M.}~\bibnamefont{Hillery}},
  \bibinfo{journal}{Phys. Rev. A} \textbf{\bibinfo{volume}{54}},
  \bibinfo{pages}{1844} (\bibinfo{year}{1996}).

\bibitem{stim1}
\bibinfo{author}{\bibfnamefont{C.}~\bibnamefont{Simon}},
  \bibinfo{author}{\bibfnamefont{G.}~\bibnamefont{Weihs}}, \bibnamefont{and}
  \bibinfo{author}{\bibfnamefont{A.}~\bibnamefont{Zeilinger}},
  \bibinfo{journal}{Phys. Rev. Lett.} \textbf{\bibinfo{volume}{84}},
  \bibinfo{pages}{2993} (\bibinfo{year}{2000}).

\bibitem{stim2}
\bibinfo{author}{\bibfnamefont{J.}~\bibnamefont{Kempe}},
  \bibinfo{author}{\bibfnamefont{C.}~\bibnamefont{Simon}}, \bibnamefont{and}
  \bibinfo{author}{\bibfnamefont{G.}~\bibnamefont{Weihs}},
  \bibinfo{journal}{Phys. Rev. A} \textbf{\bibinfo{volume}{62}},
  \bibinfo{pages}{032302} (\bibinfo{year}{2000}).

\bibitem{weidinger}
\bibinfo{author}{\bibfnamefont{M.}~\bibnamefont{Weidinger}},
  \bibinfo{author}{\bibfnamefont{B.~T.~H.} \bibnamefont{Varcoe}},
  \bibinfo{author}{\bibfnamefont{R.}~\bibnamefont{Heerlein}}, \bibnamefont{and}
  \bibinfo{author}{\bibfnamefont{H.}~\bibnamefont{Walther}},
  \bibinfo{journal}{Phys. Rev. Lett.} \textbf{\bibinfo{volume}{82}},
  \bibinfo{pages}{3795} (\bibinfo{year}{1999}).

\bibitem{bruss98b}
\bibinfo{author}{\bibfnamefont{D.}~\bibnamefont{Bruss}},
  \bibinfo{author}{\bibfnamefont{A.}~\bibnamefont{Ekert}}, \bibnamefont{and}
  \bibinfo{author}{\bibfnamefont{C.}~\bibnamefont{Macchiavello}},
  \bibinfo{journal}{Phys. Rev. Lett.} \textbf{\bibinfo{volume}{81}},
  \bibinfo{pages}{2598} (\bibinfo{year}{1998}).

\bibitem{keylwerner}
\bibinfo{author}{\bibfnamefont{M.}~\bibnamefont{Keyl}} \bibnamefont{and}
  \bibinfo{author}{\bibfnamefont{R.~F.} \bibnamefont{Werner}},
  \bibinfo{journal}{J. Math. Phys.} \textbf{\bibinfo{volume}{40}},
  \bibinfo{pages}{3283} (\bibinfo{year}{1999}).

\bibitem{qjump1}
\bibinfo{author}{\bibfnamefont{T.}~\bibnamefont{Sauter}},
  \bibinfo{author}{\bibfnamefont{W.}~\bibnamefont{Neuhauser}},
  \bibinfo{author}{\bibfnamefont{R.}~\bibnamefont{Blatt}}, \bibnamefont{and}
  \bibinfo{author}{\bibfnamefont{P.~E.} \bibnamefont{Toschek}},
  \bibinfo{journal}{Phys. Rev. Lett.} \textbf{\bibinfo{volume}{57}},
  \bibinfo{pages}{1696} (\bibinfo{year}{1986}).

\bibitem{qjump2}
\bibinfo{author}{\bibfnamefont{J.~C.} \bibnamefont{Bergquist}},
  \bibinfo{author}{\bibfnamefont{R.~G.} \bibnamefont{Hulet}},
  \bibinfo{author}{\bibfnamefont{W.~M.} \bibnamefont{Itano}}, \bibnamefont{and}
  \bibinfo{author}{\bibfnamefont{D.~J.} \bibnamefont{Wineland}},
  \bibinfo{journal}{Phys. Rev. Lett.} \textbf{\bibinfo{volume}{57}},
  \bibinfo{pages}{1699} (\bibinfo{year}{1986}).

\bibitem{plenioqjump}
\bibinfo{author}{\bibfnamefont{M.~B.} \bibnamefont{Plenio}} \bibnamefont{and}
  \bibinfo{author}{\bibfnamefont{P.~L.} \bibnamefont{Knight}},
  \bibinfo{journal}{Rev. Mod. Phys.} \textbf{\bibinfo{volume}{70}},
  \bibinfo{pages}{101} (\bibinfo{year}{1998}).

\bibitem{maitre97}
\bibinfo{author}{\bibfnamefont{X.}~\bibnamefont{Ma\^{\i}tre}},
  \bibinfo{author}{\bibfnamefont{E.}~\bibnamefont{Hagley}},
  \bibinfo{author}{\bibfnamefont{G.}~\bibnamefont{Nogues}},
  \bibinfo{author}{\bibfnamefont{C.}~\bibnamefont{Wunderlich}},
  \bibinfo{author}{\bibfnamefont{P.}~\bibnamefont{Goy}},
  \bibinfo{author}{\bibfnamefont{M.}~\bibnamefont{Brune}},
  \bibinfo{author}{\bibfnamefont{J.~M.} \bibnamefont{Raimond}},
  \bibnamefont{and} \bibinfo{author}{\bibfnamefont{S.}~\bibnamefont{Haroche}},
  \bibinfo{journal}{Phys. Rev. Lett.} \textbf{\bibinfo{volume}{79}},
  \bibinfo{pages}{769} (\bibinfo{year}{1997}).

\end{thebibliography}

\end{document}